\documentclass[letterpaper, 10 pt, conference]{ieeeconf}  % Comment this line out if you need a4paper

\IEEEoverridecommandlockouts                              % This command is only needed if 
  \let\labelindent\relax
\usepackage{amsmath} % assumes amsmath package installed
\usepackage{amssymb}  % assumes amsmath package installed

\usepackage{tabulary}
\usepackage[english]{babel}
\usepackage{blindtext}
\usepackage[makeroom]{cancel}
\usepackage{amsfonts}
\usepackage{amsthm}  %This is for proofs
\usepackage{amssymb}
\usepackage[pdftex]{graphicx}
\graphicspath{{../pdf/}{../jpeg/}}
\DeclareGraphicsExtensions{.pdf,.jpeg,.png}
\usepackage{pgf,tikz}
\usepackage{tabularx}
\usepackage{float}

\usepackage{enumitem} % To reference numbered items
\usepackage{bbm} % This is for \mathbb for lower case letters

\usepackage[margin=0.75in]{geometry}
\usepackage[strict]{changepage}

\newcommand{\vect}[1]{\boldsymbol{#1}}
\newcommand{\mat}[1]{\boldsymbol{#1}}
\newcommand{\wt}[1]{\widetilde{#1}}

\newtheorem{remark}{Remark}
\newtheorem{theorem}{Theorem}
\newtheorem{lemma}{Lemma}

\newtheorem{definition}{Definition}
\newtheorem{corollary}{Corollary}

\newtheorem{topology}{Topology Variation}

\title{\LARGE \bf
Modification of Social Dominance in Social Networks by Selective Adjustment of Interpersonal Weights
}

\author{Mengbin Ye$^{1}$, Ji Liu$^{4}$, Brian D.O. Anderson$^{1,2,3}$, Changbin Yu$^{1,2}$, and Tamer Ba\c{s}ar$^{4}$
\thanks{$^1$M. Ye, B. D.O. Anderson and C. Yu are with the Research School of Engineering, Australian National University. $^2$B. D.O. Anderson and C. Yu are with Hangzhou Dianzi University, Hangzhou, China. $^3$ B. D.O. Anderson is also with Data61-CSIRO (formerly NICTA Ltd.) in Canberra, A.C.T., Australia. 
$^4$J.~Liu and T.~Ba\c{s}ar are with the Coordinated Science
Laboratory, University of Illinois at Urbana-Champaign.
This work was supported by the Australian Research Council (ARC) under grants \mbox{DP-130103610} and \mbox{DP-160104500}, by the National Natural Science Foundation of China (grant 61375072), and by Data61-CSIRO.}% <-this % stops a space
\thanks{\hspace*{-3pt}\texttt{\{Mengbin.Ye, Brian.Anderson, Brad.Yu\}@anu.edu.au}}
\thanks{\hspace*{-3pt}\texttt{\{jiliu, basar1\}@illinois.edu}}
}

\begin{document}

\maketitle
\thispagestyle{empty}
\pagestyle{empty}

%%%%%%%%%%%%%%%%%%%%%%%%%%%%%%%%%%%%%%%%%%%%%%%%%%%%%%%%%%%%%%%%%%%%%%%%%%%%%%%%
\begin{abstract}
According to the DeGroot-Friedkin model of a social network, an individual's social power evolves as the network discusses individual opinions over a sequence of issues. Under mild assumptions on the connectivity of the network, the social power of every individual converges to a constant strictly positive value as the number of issues discussed increases. If the network has a special topology, termed ``star topology'', then all social power accumulates with the individual at the centre of the star. This paper studies the strategic introduction of new individuals and/or interpersonal relationships into a social network with star topology to \emph{reduce the social power of the centre individual.} In fact, several strategies are proposed. For each strategy, we derive necessary and sufficient conditions on the strength of the new interpersonal relationships, based on local information, which \emph{ensures that the centre individual no longer has the greatest social power within the social network.} Interpretations of these conditions show that the strategies are remarkably intuitive and that certain strategies are favourable compared to others, all of which is sociologically expected.
\end{abstract}

%%%%%%%%%%%%%%%%%%%%%%%%%%%%%%%%%%%%%%%%%%%%%%%%%%%%%%%%%%%%%%%%%%%%%%%%%%%%%%%%
\section{Introduction}\label{sec:intro}

In recent years, the systems and control community has turned to study of networked systems and multi-agent systems in the context of social sciences. Of particular interest are social networks, where groups of people interact with acquaintances through \emph{interpersonal relationships}. 

One problem of ``opinion dynamics'' has been of particular interest; how do the opinions of individuals for a given issue evolve as they discuss this issue in a social network? A recent survey on opinion dynamics is presented in \cite{friedkin2015_socialsurvey}. An important aspect of opinion dynamics is \emph{social power}, which in one sense can be considered as the weight/power/influence an individual has on the opinion discussion, relative to the weight/power/influence of the other individuals in the social network. This relativity arises due to interpersonal relationships and their strengths (which may be unidirectional). This concept is studied in the seminal works \cite{french1956_socialpower,degroot1974OpinionDynamics}. The evolution of social power is studied in \cite{friedkin2011_powerevolution}. The paper \cite{parsegov2017_multiissue} studies the case where multiple, interdependent issues are simultaneously discussed. Selecting the most influential individual in social diffusion models is studied in \cite{kempe2003_maxspread}. A social network with stubborn individuals who remain attached to their initial opinions is studied in \cite{friedkin1990_FJsocialmodel,stubborn}. The centralised DeGroot-Friedkin model for the evolution of social power is proposed and analysed in \cite{jia2015opinion_SIAM}. Distributed discrete- and continuous-time DeGroot-Friedkin models are studied in \cite{xu2015_modified_DF} and \cite{chen2017_DFdistributed} respectively. 

According to French Jr. and Snyder in \cite{french1959_leadership}, ``\emph{leadership} is the potential social influence of one part of the group over another.'' From the perspective of opinion dynamics, a leader can therefore be seen as an individual or a group of individuals that has a disproportionate amount of control over the opinion discussion process. In the context of social power, one can therefore refer to a leader/leader group as the \emph{socially dominant} individual/group of individuals. The fact that social power tends to accumulate with one individual or a subgroup of individuals in a social network is reported empirically in \cite{friedkin2011_powerevolution} and theoretically in \cite{jia2015opinion_SIAM}. This individual or subgroup is defined explicitly by the interpersonal relationships in the social network. Motivated by this concept of social dominance/leadership, and using the DeGroot-Friedkin model to describe the social network, we begin with network topologies which have a single socially dominant individual, and seek to study \emph{control strategies, including introduction of new individuals into the network and/or establishment of new interpersonal relationships, that will cause the social dominance to shift to another individual}. We now introduce the DeGroot-Friedkin model to better motivate the formal problem statement which follows in the sequel. In order to allow readers to quickly grasp the concepts of the new model and understand the motivations, in the following subsection, where possible we leave out definitions and exact mathematical results; these will be included in Section~\ref{section:background_problem}. The terms ``self-weight'',  ``individual social power'' and ``social power" will be used interchangeably.  

\subsection{The DeGroot-Friedkin Model}\label{ssec:df_model}
The discrete-time DeGroot-Friedkin model comprises a consensus model for describing the opinion dynamics (details are given below) and a mechanism for updating self-weights (the weight an individual applies to its own opinion value in the consensus process). We define $\mathcal{S} = \{0, 1, 2, \ldots\}$ to be the set of indices of sequential issues which are being discussed by the social network. For a given issue $s$, the social network discusses the issue using the discrete-time DeGroot consensus model (with constant weights throughout the discussion of the issue). At the end of the discussion (i.e. when the DeGroot model has effectively reached steady state), each individual reflects upon, and judges its impact on the discussion. This mechanism is termed reflected self-appraisal, with ``reflection'' referring to the fact that adjustments to weights are made after discussion on an issue. The individual then updates its own self-weight and discussion begins on the next issue $s+1$ (using the same consensus model but now with adjusted weights). We now explain the mathematical modelling of the mechanism for updating opinions within an issue, and the updating of self-weights from one issue to the next. 

\subsubsection{DeGroot Consensus of Opinions}
For each issue $s \in \mathcal{S}$, each individual updates its opinion $y_i(s,\cdot) \in \mathbb{R}$ at time $t+1$ as
\begin{equation}
y_i(s, t+1) = w_{ii}(s) y_i(s, t) + \sum_{j\neq i}^n w_{ij}(s) y_j(s, t)
\end{equation}
where $w_{ii}(s)$ is the self-weight individual $i$ places on its own opinion and $w_{ij}$ is the weight given by agent $i$ to the opinion of its neighbour individual $j$. As will be made apparent in the sequel, $\sum_{j = 1}^n w_{ij} = 1$, which implies that individual $i$'s new opinion value $y_i(s, t+1)$ is a convex combination of its own opinion, and the opinions of its neighbours at the current time instant. The opinion dynamics for the entire social network may be expressed as 
\begin{equation}\label{eq:opinion_network}
\vect y(s, t+1) = \mat W(s) \vect y(s, t)
\end{equation}
where $\vect y(s, t) = [y_1(s, t) \; \cdots \; y_n(s, t)]^\top$ is the vector of opinions of the $n+1$ agents in the network at time instant $t$. This model was studied in \cite{degroot1974OpinionDynamics} with $\mathcal{S} = \{0\}$ (i.e. only one issue was discussed). The dynamics of \eqref{eq:opinion_network}, and the graphical conditions required for opinions to converge, have been well studied. Next, we describe the model for the updating of $\mat W(s)$ (specifically $w_{ii}(s)$ via a reflected self-appraisal mechanism that occurs at the end of discussion of an issue $s$). For simplicity, we assume that each individual's opinion, $y_i(s,t)$, is a scalar. Kronecker products may be used if each individual's opinion state is a vector $\vect y_i \in \mathbb{R}^p, p \geq 2$.  

\subsubsection{Friedkin's Self-Appraisal Model for Determining Self-Weight} 
The Friedkin component of the model proposes a method for updating the self-weight (individual social power, self-confidence or self-esteem) of individual $i$, which is denoted by $x_i(s) = w_{ii}(s) \in [0,1]$ (the $i^{th}$ diagonal term of $\mat W(s)$) \cite{jia2015opinion_SIAM}. Define the vector $\vect x(s) = [x_1(s) \; \cdots \;  x_n (s)]^\top$ as the vector of self-weights for the individuals of the social network, with starting self-weight $0 \leq x_i(0) \leq 1$ satisfying $\sum_i x_i(0) = 1$. The self-weight vector $\vect x(s)$ is updated at the end of issue $s$ as 
\begin{equation}\label{eq:x_update}
\vect x(s+1) = \vect{\zeta}(s)
\end{equation}
where $\vect \zeta(s)^\top$ is the unique nonnegative left eigenvector of $\mat W(s)$ associated with the eigenvalue $1$, normalised such that $\vect 1_n^\top \vect \zeta(s) = 1$, see \cite{jia2015opinion_SIAM}. When individual $i$ adjusts its value of $w_{ii}(s) = x_i(s)$, it necessarily must adjust the weights $w_{ij}(s), j\neq i$ to maintain $\sum_{j=1}^n w_{ij} = 1$. The precise structure of $\mat{W}(s)$, and its properties giving rise to the existence of $\vect{\zeta}(s)^\top$, will be provided in the sequel. See Remark~\ref{rem:selfweight_update} in Section \ref{xxx} for comments on the motivation for this update mechanism. Convergence properties will be presented in the sequel, but under mild assumptions on the social network topology, it is shown that $\lim_{s\to \infty} \vect{\zeta}(s)^\top = \vect{x}^*$ where $\vect{x}^*$ is the constant vector of \emph{social power at equilibrium}.

\subsection{Contributions}
In order to simplify the problem of achieving change of a socially dominant leader, we will only consider \emph{social power at equilibrium $\vect{x}^*$} in this paper. It was shown in \cite{jia2015opinion_SIAM} that if the social network has a specific topology termed the star topology, all social power at equilibrium accumulates with a single individual $k$ as issues are sequentially discussed, in an ``autocratic configuration''. In this paper, we show that by \emph{strategic} introduction of new individuals and/or new interpersonal relationships into the social network, not only is the autocratic configuration broken but \emph{if the new relationship is sufficiently strong, other identifiable individual(s) will have social power at equilibrium \emph{greater than individual $k$}.} Specifically, we derive necessary and sufficient conditions \emph{based on local information} for the relationship strength. This is in contrast to many control strategies on networked systems which rely on global information \cite{max,consent}. In fact, a number of different strategies are considered. We also propose a strategy whereby two socially dominant individuals in separate networks can combine their networks and together remain socially dominant.

While the results are initially presented mathematically as inequalities, we provide detailed analysis and interpretation. In doing so, we show that the strategies are remarkably intuitive and precisely what one would expect when considered from a sociological context. The fact that the strategies affect the \emph{social power of individuals} in a social network which is sequentially discussing issues implies that we have developed control strategies for \emph{affecting/influencing the opinion dynamics process}.

\subsection{Paper Structure}
In Section~\ref{section:background_problem}, we provide notations, an introduction to graph theory, and convergence results for the DeGroot-Friedkin model. At the same time, a formal problem statement is given. The main results are presented in Section~\ref{sec:main_result}. Simulations are presented in Section~\ref{sec:simulations} and conclusions are drawn in Section~\ref{sec:conclusions}.

% \vspace{.2in}
% {\color{red}Ben: I have incorporated all the following literature into the intro. Ji, you may delete this paragraph when you wish.}
% {\color{blue}
% Related work:
% an early work on social power \cite{power}; a recent work on evolution of social power \cite{powerevolution}; a recent work on the case when multiple issues are discussed simultaneously \cite{multi};
% a recent survey of opinion dynamics \cite{socialsurvey}; 
% distributed discrete- and continuous-time DeGroot-Friedkin model
% \cite{acc15,xudong};

% In social diffusion models, the problem of selecting the most
% influential individuals to maximize the spread of influence is NP-hard and requires global information of the social network \cite{max}.
% }
% \vspace{.2in}

\section{Background and Formal Problem Statement}\label{section:background_problem}
% The formal problem statement appears in subsection~\ref{ssec:problem_def}. To allow proper understanding and appreciation of the motivation, we review a number of background ideas in graph theory and social network theory, where the most relevant literature is very recent.

We begin by introducing some mathematical notations used in the paper. Let $\vect 1_n$ and $\vect 0_n$ denote, respectively, the $n\times 1$ column vectors of all ones and all zeros. For a vector $\vect x\in\mathbb{R}^n$, $0\preceq\vect x$ and $0 \prec \vect x$ indicate component-wise inequalities, i.e., for all $i\in\{1,2,\ldots,n\}$, $0\leq x_i$ and $0<x_i$, respectively. Let $\Delta_n$ denote the $n$-simplex, the set which satisfies $\{\vect x\in \mathbb{R}^n : 0 \preceq \vect x, \vect 1_n^\top \vect x = 1 \}$. The canonical basis of $\mathbb{R}^n$ is given by $\mathbf{e}_1, \ldots, \mathbf{e}_n$. Define $\wt{\Delta}_n = \Delta_n \backslash \{ \mathbf{e}_1, \ldots, \mathbf{e}_n \}$ and $\text{int}(\Delta_n) = \{\vect x\in \mathbb{R}^n : 0 \prec \vect x, \vect 1_n^\top \vect x = 1 \}$. For the rest of the paper, we shall use the terms ``node'', ``agent'', and ``individual'' interchangeably. 

% An $n\times n$ matrix is called a {\em row-stochastic matrix} if its entries are all nonnegative and its row sums all equal 1. An $n\times n$ matrix is called a {\em doubly stochastic matrix} if its entries are all nonnegative and its row and column sums all equal 1.

\subsection{Graph Theory}
The interaction between individuals in a social network is modelled using a weighted directed graph, denoted as $\mathcal{G} = (\mathcal{V}, \mathcal{E})$. Each individual agent is a node in the finite, nonempty set of nodes $V = \{v_1, \ldots, v_n\}$. The set of ordered edges is $\mathcal{E} \subseteq \mathcal{V}\times \mathcal{V}$. We denote an ordered edge as $e_{ij} = (v_i, v_j) \in \mathcal{E}$, and because the graph is directed, in general $e_{ij}$ and $e_{ji}$ may not both exist. An edge $e_{ij}$ is said to be outgoing with respect to $v_i$ and incoming with respect to $v_j$. The presence of an edge $e_{ij}$ connotes that individual $j$ learns of, and takes into account, the opinion value of individual $i$  when updating its own opinion. The incoming and outgoing neighbour set of $v_i$ are respectively defined as $\mathcal{N}_i^+ = \{v_j \in \mathcal{V} : e_{ji} \in \mathcal{E}\}$ and $\mathcal{N}_i^- = \{v_j \in \mathcal{V} : e_{ij} \in \mathcal{E}\}$. The relative interaction matrix $\mat C\in\mathbb{R}^{n\times n}$ associated with $\mathcal{G}$ has nonnegative entries $c_{ij}$, termed ``relative interpersonal weights'' in \cite{jia2015opinion_SIAM}. The entries of $\mat C$ have properties such that $0 < c_{ij} \leq 1 \Leftrightarrow e_{ji} \in \mathcal{E}$ and $c_{ij} = 0$ otherwise. It is assumed that $c_{ii} = 0$ (i.e. with no self-loops), and we impose the restriction that $\sum_{j\in\mathcal{N}_i^+} c_{ij} = 1$ (i.e. that $\mat C$ is a row-stochastic matrix).

A directed path is a sequence of edges of the form $(v_{p_1}, v_{p_2}), (v_{p_2}, v_{p_3}), \ldots$ where $v_{p_i} \in \mathcal{V}, e_{ij} \in \mathcal{E}$. Node $i$ is reachable from node $j$ if there exists a directed path from $v_j$ to $v_i$. A graph is said to be strongly connected if every node is reachable from every other node. The relative interaction matrix $\mat C$ is irreducible if and only if the associated graph $\mathcal{G}$ is strongly connected ($\mat C$ is known in some literature as the weighted adjacency matrix). If $\mat C$ is irreducible then it has a unique (up to a scaling) left eigenvector $\vect{\gamma}^\top$, with all entries strictly positive, associated with the eigenvalue 1 (Perron-Frobenius Theorem, see \cite{godsil2001algebraic}). Henceforth, we shall call this left eigenvector $\vect{\gamma}^\top$ the \emph{dominant left eigenvector of $\mat C$} and assume that $\vect{\gamma}^\top$ has been normalised such that it has the property $\vect{\gamma}^\top\vect 1_n = 1$.

\subsection{Convergence Results for the DeGroot-Friedkin Model}\label{xxx}
We now provide additional, specific details on the model. For a given issue, the influence matrix $\mat W(s)$ is defined as follows 
\begin{equation}\label{eq:W_matrix}
\mat W(s) = \mat X(s) + (\mat I_n - \mat X(s))\mat C
\end{equation}
where $\mat C$ is the relative interaction matrix associated with the graph $\mathcal{G}$, and the matrix $\mat X(s) \doteq diag[\vect x(s)]$. From the fact that $\mat C$ is row-stochastic with zero diagonal entries, \eqref{eq:W_matrix} implies that $\mat W(s)$ is a row-stochastic matrix. Note that $\vect{\zeta}(s)^\top \vect{1}_n = 1$ implies that $\vect x(s) \in \Delta_n$ for all $s$. From \eqref{eq:W_matrix} and the fact that $\mat{C}$ is constant, it is apparent that by adjusting $w_{ii}(s+1) = \zeta_i(s)$, individual $i$ also scales $w_{ij}(s+1), j \neq i$ by $(1-w_{ii}(s+1))$ in order to maintain the row-stochastic property of $\mat{W}(s)$.

\begin{remark}[Social Power]\label{rem:selfweight_update}
The precise motivation behind using \eqref{eq:x_update} as the updating model for $\vect{x}(s)$ is detailed in \cite{jia2015opinion_SIAM}, but we provide a brief overview here in the interest of making this paper self-contained. The definition of $\mat{W}$ in \eqref{eq:W_matrix} ensures that, for any given $s$, there holds $\lim_{t \to \infty} y(s,t) = (\vect{\zeta}(s)^\top \vect y(s,0))\vect{1}_n$. In other words, for any given issue $s$, the opinions of every individual in the social network reach a consensus value $\vect{\zeta}(s)^\top \vect y(s,0)$ equal to a convex combination of their initial opinion values $\vect y(s,0)$. The elements of $\vect{w}(s)^\top$ are the convex combination coefficients, i.e. $\zeta_{i}(s)$ represents precisely the amount of weight/power that individual $i$ had on the opinion discussion for issue $s$. For a given issue $s$, $\zeta_i(s)$ is a manifestation of individual $i$'s social power in the social network, as it is in effect the ability of individual $i$ to control the outcome of a discussion \cite{cartwright1959social_book}. The reflected self-appraisal mechanism therefore describes an individual $1)$ observing how much power it had on the discussion of issue $s$ (the nonnegative quantity $\zeta_{i}(s)$) and, $2)$ for the following issue $s+1$, adjusting its self-weight to be equal to this power, i.e. $x_i(s+1) = w_{ii}(s+1) = \zeta_i(s)$. As can be observed from \eqref{eq:W_matrix} and because $\mat{C}$ is constant, adjusting $w_{ii}(s+1)$ also adjusts the interpersonal weights $w_{ij}(s+1)$.
\end{remark}

It is shown in [Lemma 2.2, \cite{jia2015opinion_SIAM}] that the system \eqref{eq:x_update} describing the update of self-weights, is equivalent to 
\begin{equation}\label{eq:DF_system}
\vect x(s+1) = \vect F(\vect x(s))
\end{equation}
where the nonlinear vector-valued function $\vect F(\vect x(s))$ is defined as
\begin{align}\label{eq:map_F_DF}
\vect F( \vect x(s) ) =   \begin{cases} 
   \mathbf e_i & \hspace*{-6pt} \text{if } x_i(s) = \mathbf e_i, \text{for any } i \\ \\
   \alpha (\vect x(s)) \begin{bmatrix} \frac{c_1}{1-x_1(s)} \\ \vdots \\ \frac{c_n}{1-x_n(s)} \end{bmatrix}       & \text{otherwise }
  \end{cases}
\end{align}
with $\alpha(\vect x(s)) = 1/\sum_{i=1}^n \frac{c_i}{1- x_i(s)}$. Much of this paper will deal with scenarios where the underlying graph has a star topology or its variants, the definition and relevance of which are now given.
\begin{definition}[Star topology]\label{def:star}
A strongly connected graph\footnote{While it is indeed possible to have a star graph that is not strongly connected, this paper similarly to \cite{jia2015opinion_SIAM} deals only with graphs which are strongly connected.} $\mathcal{G}$ is said to have star topology if there exists a node $i$, which is called the centre node, such that every edge of $\mathcal{G}$ is either to or from node $i$
\end{definition}
Note that the irreducibility of $\mat C$ implies that the star topology must include edges in both directions between the centre node $v_i$ and every other node $v_j, j \neq i$. We now provide a lemma and a theorem regarding the convergence of $\vect{F}(\vect{x}(s))$ as $s\to \infty$. 

\begin{lemma}[Lemma 3.2, \cite{jia2015opinion_SIAM}]\label{lem:star}
Suppose that $n \geq 3$, and suppose further that $\mathcal{G}$ has star topology, which without loss of generality has centre node $v_1$. Let $\mat{C}$ be the row-stochastic and irreducible adjacency matrix, with zero diagonal entries, associated with $\mathcal{G}$. Then for all initial conditions $\vect{x}(0) \in \wt{\Delta}_n$, the self-weights $\vect x(s)$ converge to the fixed point $\vect{x}^* = \mathbf{e}_1$ as $s \to \infty$.
\end{lemma}

\begin{theorem}[Theorem 4.1, \cite{jia2015opinion_SIAM}]\label{thm:DFmain}
For $n\geq 3$, consider the DeGroot-Friedkin dynamical system \eqref{eq:DF_system} with a relative interaction matrix $\mat C$ that is row-stochastic, irreducible, and has zero diagonal entries. Assume that the digraph $\mathcal{G}$ associated with $\mat C$ does not have star topology and define $\vect\gamma^\top$ as the dominant left eigenvector of $\mat C$. Then,
\begin{enumerate}[label=(\roman*)]
\item \label{prty:thm_DFmain01} For all initial conditions $\vect x(1) \in \wt{\Delta}_n $, the self-weights $\vect x(s)$ converge to $\vect x^*$ as $s\to\infty$. Here, $\vect x^* \in \wt{\Delta}_n $ is the unique fixed point satisfying $\vect x^* = \vect F(\vect x^*)$. 
\item \label{prty:thm_DFmain02} There holds $x^*_i < x^*_j$ if and only if $\gamma_i < \gamma_j$, for any $i,j$, where $\gamma_i$ is the $i^{th}$ entry of the dominant left eigenvector $\vect\gamma$. There holds $x^*_i = x^*_j$ if and only if $\gamma_i = \gamma_j$. 
\item \label{prty:thm_DFmain03} The unique fixed point $\vect x^*$ is determined only by $\vect\gamma^\top$, and is independent of the initial conditions. 
\end{enumerate}
\end{theorem}

An interpretation of Lemma~\ref{lem:star} and Theorem~\ref{thm:DFmain} is given in below in Remark~\ref{rem:autocratic}.

\subsection{Formal Problem Statement}\label{ssec:problem_def}
In this paper, we investigate \emph{how additional nodes and/or edges strategically connected to a star topology can change the social power at equilibrium, $\vect{x}^*$}. To that end, we begin first by providing definitions which will aid in describing our problem and discussing results obtained. Moreover, we are interested in comparing the social power of individuals within the network at equilibrium, i.e. when $s\to \infty$. We will therefore refer to the equilibrium value $x_i^*$ as the social power of individual $i$ when there is no ambiguity (as opposed to the evolving $x_i(s)$ when $s < \infty$). 

To simplify the problem, we do not study the evolution of the opinions $\vect y(s,t)$. Under the assumption that $\mat C$ is irreducible, it is shown in \cite{jia2015opinion_SIAM} that, for any issue $s$, the opinions always converge as $\lim_{t\to \infty} \vect{y}(s,t) = (\zeta(s)^\top \vect{y}(s,0))\vect{1}_n$. As discussed above, we are interested in individual social power of the network.

\begin{definition}[Autocratic Network]\label{def:autocrat}
A social network is said to be an autocratic configuration, with node $v_i$ being the autocrat, if $\vect x(s)  = \mathbf{e}_i$. 
\end{definition}

\begin{definition}[Social dominance/leadership]\label{def:leader}
Node $v_i$ is said to be the socially dominant/leader node in the network if $x_i^* > x_j^*$ for all $j \neq i$. In other words, at equilibrium,  the social power of individual $i$ is greater than the social power of any other individual in the social network.
\end{definition}

\begin{remark}[Autocratic tendency]\label{rem:autocratic}
Lemma~\ref{lem:star} has an important social connotation. One can consider $x_i(0)$ as individual $i$'s \emph{estimate} of its social power when the social network is first formed, before any issue discussion. For any initial estimate $\vect{x}(0) \in \wt{\Delta}_n$ (that is, no individual $i$ believes $x_i(0) = 1$), the star topology network tends to an autocratic configuration at equilibrium, $\vect x^* = \mathbf{e}_1$. This implies that, for the first few issues, opinion discussion will occur with everyone contributing to the final consensus value. However, as more issues are discussed, the centre individual increasingly guides the outcome of discussions until, for $s = \infty$, only the centre individual's opinion value matters.
\end{remark}

\begin{remark}\label{rem:trust}
In \cite{jia2015opinion_SIAM}, the constant entries $c_{ij}$ of $\mat{C}$ are termed ``relative interpersonal weights'', and we will keep with this terminology. However, one can also consider $c_{ij}$ as the amount of ``trust'' individual $i$ has for individual $j$ or the strength of ``influence'' individual $j$ has on individual $i$. In other words, $c_{ij}$ captures the strength of a unidirectional relationship (unidirectional since $c_{ij} \neq c_{ji}$ in general). 
\end{remark}

\begin{figure*}
\begin{minipage}{0.32\linewidth}
\begin{center}
\resizebox{0.75\columnwidth}{!}{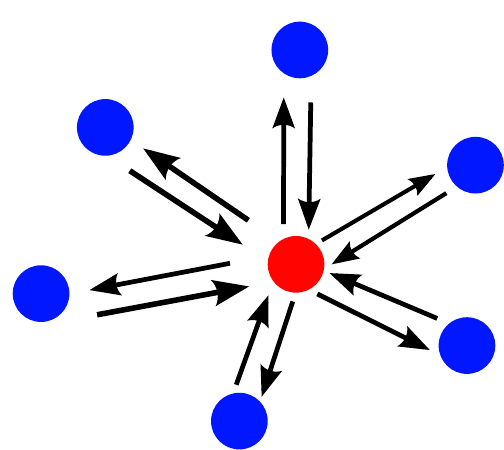}
\caption{Star Topology with red Centre Node $v_1$, and blue subject nodes, $n = 7$.}
\label{fig:star_top}
\end{center}
\end{minipage}
\hfill
\begin{minipage}{0.32\linewidth}
\begin{center}
\resizebox{0.9\columnwidth}{!}{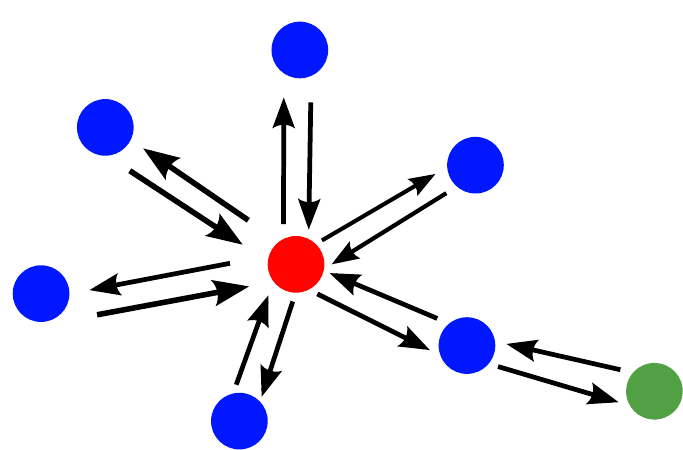}
\caption{Topology Variation~\ref{top:single} (Single Attacker) with $n = 8$, attacker nodes are green. }
\label{fig:single_attack}
\end{center}
\end{minipage}
\hfill
\begin{minipage}{0.32\linewidth}
\begin{center}
\resizebox{1.1\columnwidth}{!}{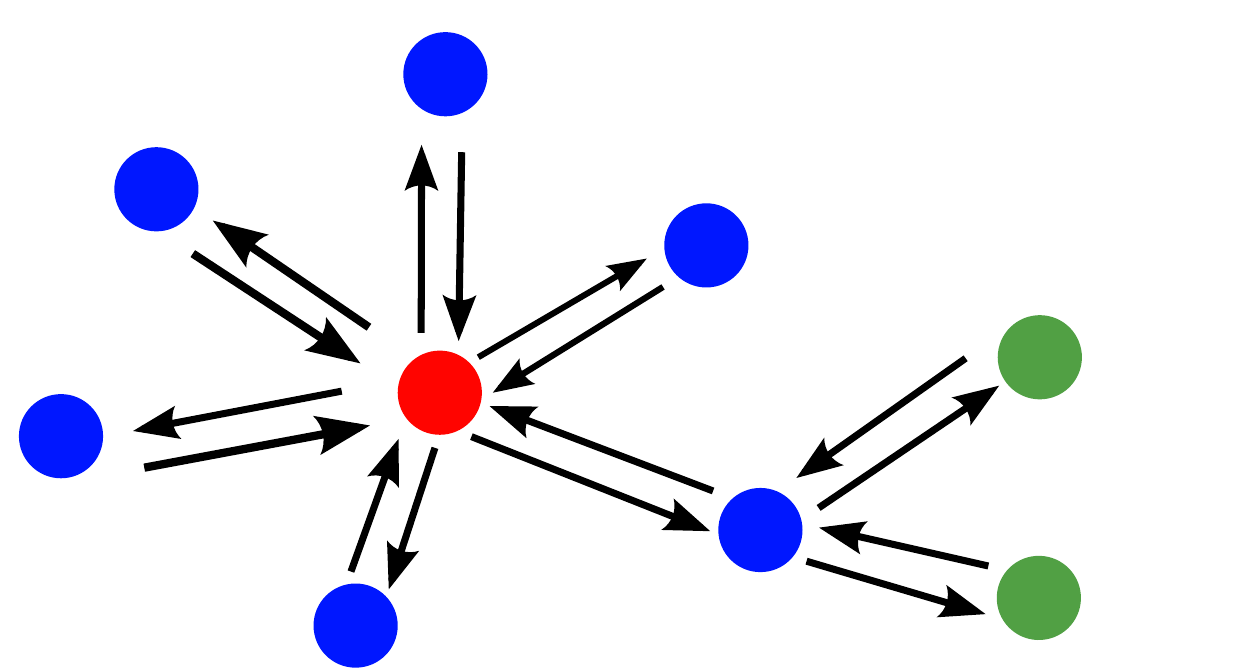}
\caption{Topology Variation~\ref{top:two_attack} (Coordinated Double Attacker) with $n = 9$, attacker nodes are green.}
\label{fig:two_attack}
\end{center}
\end{minipage}
\end{figure*}

\begin{figure*}
\begin{minipage}{0.32\linewidth}
\begin{center}
\resizebox{1\columnwidth}{!}{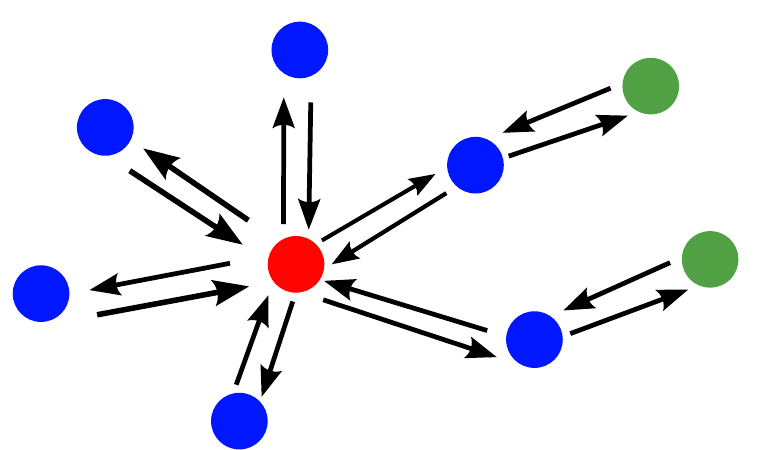}
\caption{Topology Variation~\ref{top:two_attack_two_sub} (Uncoordinated Double Attacker) with $n = 9$.}
\label{fig:two_attack_two_sub}
\end{center}
\end{minipage}
\hfill
\begin{minipage}{0.32\linewidth}
\begin{center}
\resizebox{1\columnwidth}{!}{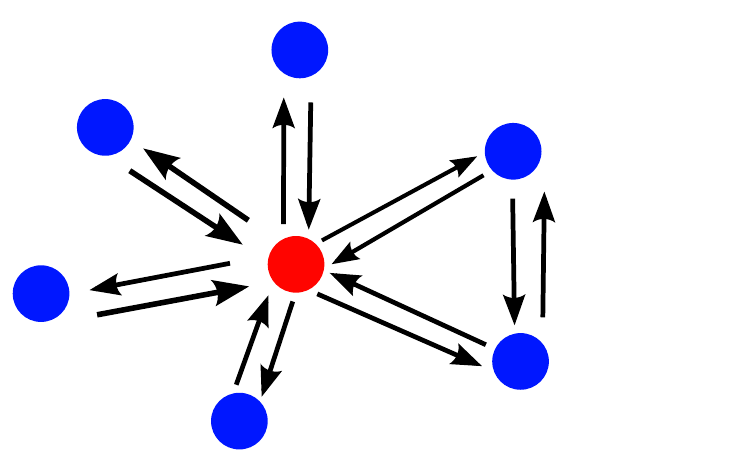}
\caption{Topology Variation~\ref{top:no_attack} (Two Dissenting Subjects) with $n = 7$.}
\label{fig:no_attack}
\end{center}
\end{minipage}
\hfill
\begin{minipage}{0.32\linewidth}
\begin{center}
\resizebox{1\columnwidth}{!}{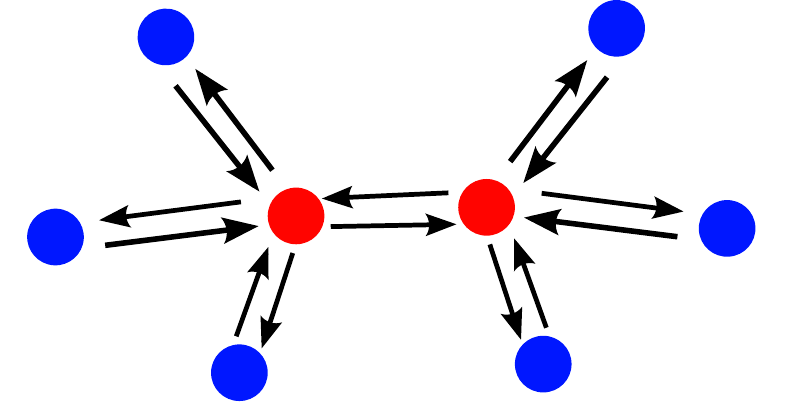}
\caption{Topology Variation~\ref{top:leader_group} (Leadership Group) with $n = m = 4$.}
\label{fig:leader_group}
\end{center}
\end{minipage}
\end{figure*}

For a given graph $\mathcal{G}$ with star topology, with centre node $v_1$, let us call the other nodes \emph{subject nodes} in the sense that they are subjects to the autocrat centre node. In Fig.~\ref{fig:star_top}, these are nodes $v_i, i = 2, ..., 7$. We are going to study how the autocracy can be disrupted by introduction of a perturbation to the star graph. This leads us to define a new type of node. An \emph{attacker node} is a node $v_j$ which forms edges $e_{ji}, e_{ij}$ with some node $v_i$, $i \neq 1$, i.e. a subject node. In doing so, we modify the graph $\mathcal{G}$ to become $\bar{\mathcal{G}}$ which is no longer a star. In Fig.~\ref{fig:single_attack}, node $v_8$ is the attacker node, forming edges with node $v_7$. We call node $v_j$ an attacker node because, as will become apparent in the sequel, the weights $c_{ji}$ and $c_{ij}$ determine the social power $x_1^*$ of the autocrat node $v_1$. In other words, $v_j$ attacks the social dominance of $v_1$. Note that two edges, $e_{ji}, e_{ij}$ must be formed to ensure that $\bar{\mathcal{G}}$ remains strongly connected. Actually, there are a number of interesting ways to attack the social dominance of $v_1$, and we list some of the most important/fundamental methods. For each topology variation we list below, we provide an example in the Figures~\ref{fig:single_attack}-\ref{fig:leader_group}.

\begin{topology}[Single Attack]\label{top:single}
Suppose that $n\geq 4$. Suppose further that $\mathcal{G}$ has star topology, with $v_1$ being the centre node, and with $n-2$ subject nodes, $v_i, i = 2, ..., n-1$. A single attacker node $v_n$ attaches to subject node $v_{n-1}$ by forming edges $e_{n-1,n}$ and $e_{n,n-1}$. This forms the modified graph $\bar{\mathcal{G}}$
\end{topology}

\begin{topology}[Coordinated Double Attack]\label{top:two_attack}
Suppose that $n\geq 5$. Suppose further that $\mathcal{G}$ has star topology, with $v_1$ being the centre node, and with $n-3$ subject nodes, $v_i, i = 2, ..., n-2$. Two attacker nodes $v_{n-1}$ and $v_n$ attach to subject node $v_{n-2}$ by forming the set of edges $\{e_{n-2,n-1}, e_{n-1,n-2}, e_{n-2,n}, e_{n,n-2}\}$. This forms $\bar{\mathcal{G}}$.
\end{topology}

\begin{topology}[Uncoordinated Double Attack]\label{top:two_attack_two_sub}
Suppose that $n \geq 5$. Suppose further that $\mathcal{G}$ has star topology, with $v_1$ being the centre node, and with $n-3$ subject nodes, $v_i, i = 2, ..., n-2$. One attacker node $v_{n-1}$ attaches to subject node $v_{n-3}$ with edges $e_{n-3,n-1}, e_{n-1,n-3}$. A second attacker node $v_{n}$ attaches to subject node $v_{n-2}$ with edges $e_{n-2,n}, e_{n,n-2}$. This forms $\bar{\mathcal{G}}$.
\end{topology}

\begin{topology}[Two Dissenting Subjects]\label{top:no_attack}
Suppose that $n \geq 4$. Suppose further that $\mathcal{G}$ has star topology, with $v_1$ being the centre node, and with $n-1$ subject nodes, $v_i, i = 2, ..., n$. There are no attacker nodes. Subject nodes $v_{n-1}$ and $v_n$ form edges $e_{n,n-1}, e_{n-1,n}$, forming $\bar{\mathcal{G}}$.
\end{topology}

The following topology variation is motivated by the concept of a \emph{leadership group} where two leaders exist, and seek to maintain their collective social dominance.

\begin{topology}[Leadership group]\label{top:leader_group}
Suppose that $\mathcal{G}_1$ and $\mathcal{G}_2$ respectively have $n \geq 3$ and $m\geq 3$ nodes, with node set $\mathcal{V}_1 = \{1, ..., n\}$ and $\mathcal{V}_2 = \{ n+1, ..., n+m\}$ respectively. Both $\mathcal{G}_1$ and $\mathcal{G}_2$ have star topology; the centre nodes for $\mathcal{G}_1$ and $\mathcal{G}_2$ are $v_1$ and $v_{n+1}$ respectively. Let $\bar{\mathcal{G}}$ be the graph formed by merging $\mathcal{G}_1$ and $\mathcal{G}_2$ by insertion of the edges $e_{1,n+1}$ and $e_{n+1,1}$. Nodes $v_1, v_{n+1}$ form a \emph{leadership group} with subjects $v_2, ..., v_n, v_{n+2}, ..., v_{n+m}$.
\end{topology}

In the next section, we investigate the above topological variations of the star graph. Note that Topology Variations~\ref{top:single}-\ref{top:leader_group} have modified graphs $\bar{\mathcal{G}}$ which \emph{do not have star topology}. From the properties of $\vect{F}(\vect{x}(s))$ established in \cite{jia2015opinion_SIAM} and detailed in Lemma~\ref{lem:star} and Theorem~\ref{thm:DFmain}, it immediately follows that $x_1^* < 1$ for all Topology Variations. In other words, $v_1$ is no longer the autocrat but if the perturbation from star topology (caused by the new edges) is small, one expects that $v_1$ remains \emph{socially dominant}. What will we show is that if the interpersonal weights associated with these new edges exceed a given threshold, the socially dominant node changes from $v_1$ to some other node. It is worth emphasising at this stage that, in Variations~\ref{top:single}-\ref{top:two_attack_two_sub}, it is useless for an attacker node $v_n$ to attach to the centre node $v_1$ instead of a subject node; the topology remains a strongly connected star, and there is no change in the autocratic nature of $v_1$'s social dominance.

Note that when new edges are introduced, we assume each individual $i$ adjusts its weights $c_{ij}$ to ensure that the new $\mat{C}$ is row-stochastic. Take Topology Variation~\ref{top:leader_group} as an example. Separately, the relative interaction matrix $\mat{C}_1$ (respectively $\mat{C}_2$) associated with $\mathcal{G}_1$ (respectively $\mathcal{G}_2$) is assumed to be row-stochastic. The relative interaction matrix $\bar{\mat C}$ associated with $\bar{\mathcal{G}}$ is also implicitly assumed to be row-stochastic with zero diagonal. That is, we assume that after the addition of edges $e_{1,n+1}$ and $e_{n+1,1}$, adjustments are made to the original weights $c_{1,j}$ and $c_{n+1,k}$ to ensure $\bar{\mat{C}}$ is row-stochastic.

\begin{remark}[Ordering of Social Power]
Although Theorem~\ref{thm:DFmain} states that $\vect{x}^*$ is uniquely determined by $\vect{\gamma}^\top$, there are no results available which allow one to analytically compute the \emph{value} of $\vect{x}^*$ given $\vect{\gamma}^\top$. What is available is Statement~\ref{prty:thm_DFmain02} of Theorem~\ref{thm:DFmain}, which states that the ordering $\vect{x}_i^*$ is consistent with the ordering of $\gamma_i$. In this paper, we are therefore interested in the ordering of individual social power, as opposed to the precise values of social power. This is reflected in Definition~\ref{def:leader}.
\end{remark}

\section{Main Results}\label{sec:main_result}
In order to maintain the flow of this paper, and to place focus on discussion of the \emph{social connotations} of each result, we place all proofs in the appendix. We firstly present theorems and corollaries for each topological variation, and then discuss their social implications.

\subsection{Topology Variation~\ref{top:single}: Single Attacker}
Now, firstly consider Topology Variation~\ref{top:single}. The relative interaction matrix $\mat C(\beta)$ associated with $\bar{\mathcal{G}}$ can be expressed as 
\begin{equation}\label{eq:C_singleattack}
\mat C(\beta) = 
\begin{bmatrix}
0 & c_{12} & c_{13} & \hdots & c_{1,n-1} & 0 \\
1 & 0 & 0 & \hdots & 0 & 0 \\
1 & 0 & 0 & \hdots & 0 & 0 \\
\vdots & \vdots & \vdots  & \ddots & \vdots & \vdots \\
 1-\beta &  0 & 0 & \hdots & 0 & \beta \\
  0 & 0 & 0 & \hdots & 1 & 0
\end{bmatrix}
\end{equation}
where $\beta = c_{n-1,n} \in (0,1)$ is the influence exerted by the attacker node $v_n$ on subject node $v_{n-1}$. The following theorem details how the social power of each individual changes as $\beta$ changes.

\begin{theorem}[Single Attack]\label{thm:single}
For a social network with Topology Variation~\ref{top:single}, with initial conditions $\vect{x}(0) \in \wt{\Delta}_n$, and described by the DeGroot-Friedkin model, the following statements are true:
\begin{enumerate}[label=(\roman*)]
\item \label{prty:thm_single01} For all values of $\beta \in (0,1)$, there holds $x_i^* < x_1^*$, for all $i \neq 1, n-1, n$ and $x_n^* < x_{n-1}^*$
\item \label{prty:thm_single02} There holds 1) $x_1^* > x_i^*, \forall\, i \neq 1$ if and only if $\beta <  1- c_{1,n-1} = \sum_{i = 2}^{n-2} c_{1,i}$, 
or 2) $x_{n-1}^* > x_i^*, \forall\, i \neq n-1$
if and only if $\beta >  1- c_{1,n-1}$. There holds $x_1^* = x_{n-1}^* > x_i^*, \forall\, i \neq 1,n-1$ if and only if $\beta_1 = 1- c_{1,n-1}$
\item \label{prty:thm_single03} There holds $x_n^* > x_1^*$ if and only if $\beta > 1/(1+c_{1,n-1})$.
\end{enumerate}
\end{theorem}

\begin{corollary}[Generalised Placement of Single Attacking Node]\label{cor:single}
Suppose that instead of attaching to subject node $v_{n-1}$, attacker node $v_n$ can attach to any subject node $v_i, i \in \{2, ..., n-1\}$ by forming edges $e_{n,i}, e_{i,n}$. The lower bound on $\beta = c_{n,i}$ required to have $x_{n-1}^* > x_1^*$ is minimised if $v_n$ attaches to $v_k$ where $k = \text{argmax}_{j\in \{2, ..., n-1\}} \; c_{1,j}$.
\end{corollary}

The above mathematical results can be interpreted in the following social context. From Statement~\ref{prty:thm_single01}, we conclude that individuals $2$ to $n-2$, i.e. subject nodes $v_i, i\in \{2, .., n-2\}$ will never have greater social power at equilibrium, $x_i^*$ than the centre individual $v_1$ with $x_1^*$, regardless of how $\beta$ changes. In addition, the attacker node will never have greater social power than the subject node $v_{n-1}$ which it is attached to.

Recall from Remark~\ref{rem:trust} that $c_{ij}$ can be considered the trust level accorded to individual $j$ by individual $i$. Then according to Statement~\ref{prty:thm_single02}, centre individual $v_1$ \emph{remains the socially dominant individual in the social network} only if subject $v_{n-1}$ trusts attacker $v_n$ less than the total sum of trust accorded to subjects $v_i, i\in \{2, ..., n-2\}$ by centre node $v_1$. \emph{In order to become socially dominant, and to undermine the authority of the centre node $v_1$, individual $v_{n-1}$ must trust the attacker $v_n$}. 

Lastly, Statement~\ref{prty:thm_single03} reveals that the attacker can also obtain social power greater the centre individual $v_1$ if $\beta$, i.e. the trust accorded to the attacker $v_n$ by subject node $v_{n-1}$, is sufficiently large. We therefore conclude that the leadership/social dominance within a social network with Topology Variation~\ref{top:single} can be shifted from the original leader (centre node) to a subject via introduction of a single attacker and strengthening of the newly formed interpersonal relationships.

Corollary~\ref{cor:single} delivers an intuitive and powerful, socially relevant result. It states that the single attacker node $v_n$ should seek to form an interpersonal relationship with \emph{the subject node $v_k$ that centre node $v_1$ trusts the most}. This will minimise the required amount of trust subject $v_k$ accords attacker $v_n$ before centre node $v_1$ loses social dominance.

\subsection{Topology Variation~\ref{top:two_attack}: Coordinated Double Attack}

Consider now Topology Variation~\ref{top:two_attack}. Firstly, define $\beta_{1} = c_{n-2,n-1} \in (0,1)$ and $\beta_{2} = c_{n-2,n} = (0,1)$ as the two adjustable interpersonal weights. Note that because $\mat{C}$ is assumed to be row-stochastic, it is implied that $\beta_1 + \beta_2 + c_{n-2,1} = 1$ which in turn implies $\beta_1 + \beta_2 < 1$ because $c_{n-2,1} > 0$.  We omit displaying the exact form of $\mat{C}(\beta_1, \beta_2)$ due to spatial limitations. 
\begin{theorem}[Coordinated Double Attack]\label{thm:two_attack}
For a social network with Topology Variation~\ref{top:two_attack}, with initial conditions $\vect{x}(0) \in \wt{\Delta}_n$, and described by the DeGroot-Friedkin model, the following statements are true:
\begin{enumerate}[label=(\roman*)]
\item \label{prty:thm_two01} For all $\beta_1, \beta_2 \in (0,1)$, there holds $x_i^* < x_1^*$ for all $i \neq 1, n-2, n-1, n$, and $x_n^*, x_{n-1}^* < x_{n-2}^*$.
\item \label{prty:thm_two02} There holds 1) $x_1^* > x_i^*, \forall, i \neq 1$ if and only if $\beta_1+\beta_2 <  1- c_{1,n-2} = \sum_{i = 2}^{n-3} c_{1,i}$, or 2) $x_{n-2}^* > x_i^*, \forall, i \neq n-2$
if and only if $\beta_1+\beta_2 >  1- c_{1,n-2}$. There holds $x_1^* = x_{n-2}^*$ if and only if $\beta_1+\beta_2 =  1- c_{1,n-2}$.
\item \label{prty:thm_two03} There holds $x_n^* > x_1^*$ (respectively $x_{n-1}^* > x_1^*$) if and only if $\beta_2 > (1-\beta_1)/(1+c_{1,n-2})$ (respectively $\beta_1 > (1-\beta_2)/(1+c_{1,n-2})$ ).
\item \label{prty:thm_two04} There holds $x_{n-1}^* < x_{n}^*$ or $x_{n-1}^* > x_{n}^*$ if and only if $\beta_1 < \beta_2$ or $\beta_1 > \beta_2$ respectively. If $\beta_1 = \beta_2$, then $x_{n-1}^* = x_{n}^*$.
\end{enumerate}
\end{theorem}

\begin{corollary}[Generalised Placement of Coordinated Double Attack]\label{cor:two_attack}
Suppose that instead of attaching to subject node $v_{n-2}$, attacker nodes $v_{n-1}, v_{n}$ can attach to any subject node $v_i, i \in \{2, ..., n-2\}$ by forming the set of edges $\{ e_{n-1,i}, e_{i,n-1}, e_{n,i}, e_{i,n} \}$. The lower bound on $\beta_1+ \beta_2 = c_{n-1,i} + c_{n,i}$ required to have $x_{n-2}^* > x_1^*$ is minimised if $v_{n-1}$ and $v_n$ attach to $v_k$ where $k = \text{argmax}_{j\in \{2, ..., n-2\}} \; c_{1,j}$.
\end{corollary}

Due to spatial limitations, we discuss social implications of Theorem~\ref{thm:two_attack} only if the conclusions differ significantly from the discussion in the previous subsection. 

The key result is Statement~\ref{prty:thm_two02}, which indicates that the \emph{combined trust} given to attackers $v_{n-1}$ and $v_n$ by subject node $v_{n-2}$ must exceed the \emph{combined trust} given to subjects $v_2, ..., v_{n-3}$ by centre node $v_1$, in order for centre node $v_1$ to lose social dominance (and thus subject $v_{n-2}$ becomes the socially dominant individual). It is most interesting to note that \emph{it is only the sum} of the trust/influence $\beta_1 + \beta_2$ that is relevant, and \emph{there is no requirement on the individual magnitudes of $\beta_1, \beta_2$}. 

Regarding Statement~\ref{prty:thm_two03}, we observe that the inequality, which if satisfied ensures that attacker $v_n$ has social power greater than centre $v_1$, is a function of $\beta_1, \beta_2$ and $c_{1,n-2}$. As detailed in the proof in Appendix~\ref{app:pf_twoattack}, there always exists a $\beta_1, \beta_2$ satisfying $\beta_1 + \beta_2 < 1$ which ensures both attacker nodes $v_{n-1}, v_{n}$ have social power greater than the centre $v_1$.

\subsection{Topology Variation~\ref{top:two_attack_two_sub}: Uncoordinated Double Attack}
Define $\beta_1 = c_{n-3,n-1} \in (0,1)$ and $\beta_2 = c_{n-2,n} \in (0,1)$.

\begin{theorem}\label{thm:twotwo}
For a social network with Topology Variation~\ref{top:two_attack_two_sub}, with initial conditions $\vect{x}(0) \in \wt{\Delta}_n$, and described by the DeGroot-Friedkin model, the following statements are true:
\begin{enumerate}[label=(\roman*)]
\item \label{prty:thm_twotwo01} For all values of $\beta_1,\beta_2 \in (0,1)$, there holds $x_i^* < x_1^*$ for all $i \in\{2,\ldots,n-4\}$, and $x_{n-1}^* < x_{n-3}^*$ and $x_{n}^* < x_{n-2}^*$.
\item \label{prty:thm_twotwo02} There holds $x_1^* > x_i^*$ for all $i \neq 1$ if and only if $\beta_1 <  1- c_{1,n-3}$
and $\beta_2 < 1-c_{1,n-2}$. If $\beta_1 >  1- c_{1,n-3}$ (respectively $\beta_2 > 1 - c_{1,n-2}$), then $x_{n-3}^* > x_1^*$ (respectively $x_{n-2}^* > x_1^*$).
\item \label{prty:thm_twotwo03} For $i\in\{1,2\}$, there holds $x_{n-2+i}^* > x_1^*$ if and only if $\beta_i > 1/(1+c_{1,n-4+i})$.
\item \label{prty:thm_twotwo04} There holds $x_{n-3}^* > x_{n-2}^*$ if and only if $\frac{1-\beta_2}{1-\beta_1}>\frac{c_{1,n-2} }{c_{1,n-3}}$. Equivalently, $x_{n-3}^* > x_{n-2}^*$ if and only if $\frac{c_{1,n-3}}{c_{n-3,1}}>\frac{c_{1,n-2} }{c_{n-2,1}}$
\end{enumerate}
\end{theorem}

The most interesting conclusion drawn from Theorem~\ref{thm:twotwo} is when we compare to Theorem~\ref{thm:two_attack} which concerns Topology Variation~\ref{top:two_attack}. With Topology Variation~\ref{top:two_attack}, for the centre individual $v_1$ to lose its social dominance we require \emph{the sum of the trust values $\beta_1 + \beta_2$} to exceed a lower bound, and there are no separate lower bounding inequalities for $\beta_1$ or $\beta_2$. With Topology Variation~\ref{top:two_attack_two_sub}, centre individual $v_1$ loses social dominance if and only if either $\beta_1$ or $\beta_2$ exceed their respective lower bounding inequalities. Importantly, these two lower bounding inequalities \emph{are independent of each other}. \emph{This clearly points to the fact that a \textbf{coordinated attack} on the social dominance of the centre node is more desirable, an idea which is socially intuitive}.

From Statement \ref{prty:thm_twotwo03}, both attacker nodes have larger social power than the centre node if and only if 
$\beta_1 > 1/(1+c_{1,n-3})$ and $\beta_2 > 1/(1+c_{1,n-2})$, 
which implies that $\beta_1+\beta_2 > 1/(1+c_{1,n-3})+1/(1+c_{1,n-2})$. From Statement \ref{prty:thm_two03} Theorem \ref{thm:two_attack}, with Topology Variation \ref{top:two_attack}, both attacker nodes have larger social power than the centre node if and only if $\beta_2 > (1-\beta_1)/(1+c_{1,n-2})$ and $\beta_1 > (1-\beta_2)/(1+c_{1,n-2})$, which implies that $\beta_1+\beta_2 > 2/(2+c_{1,n-2})$. Since both $1+c_{1,n-2}$ and $1+c_{1,n-3}$ are smaller than $2+c_{1,n-2}$, it follows that 
$1/(1+c_{1,n-3})+1/(1+c_{1,n-2})> 2/(2+c_{1,n-2})$, which implies that \emph{a \textbf{coordinated attack} on the social dominance of the centre node is also more efficient for the attackers}. 

\subsection{Topology Variation~\ref{top:no_attack}: Two Dissenting Subjects}

Topology Variation~\ref{top:no_attack} is different to the ones studied above in the sense that there are no attacker nodes. Instead, one can consider this variation as one where two subjects form a relationship in \emph{dissent} from the leader. Firstly, let $\beta_1 = c_{n-1,n} \in (0,1)$ and $\beta_2 = c_{n,n-1} \in (0,1)$. Analysis yields the following result.

\begin{theorem}[Two Dissenting Subjects]\label{thm:no_attack}
For a social network with Topology Variation~\ref{top:no_attack}, with initial conditions $\vect{x}(s) \in \wt{\Delta}_n$, and described by the DeGroot-Friedkin model, the following statements are true: 
\begin{enumerate}[label=(\roman*)]
\item \label{prty:thm_no01} For all $\beta_1, \beta_2 \in (0,1)$, there holds $x_i^* < x_1^*$ for all $i \neq 1, n-1, n$.
\item \label{prty:thm_no02} There holds $x_n^* > x_1^*$ if and only if $\beta_1 > (1-c_{1,n})/(c_{1,n-1}+\beta_2)$ with $\beta_1 \in (0,1)$. There exists such a $\beta_1 \in (0,1)$ only if $\beta_2 > \sum_{i = 2}^{n-2} c_{1,i}$. 
%\item \label{prty:thm_no03} There holds $x_n^* > x_1^*$ if $\beta_2 > (1-c_{1,n} - \beta_1 c_{1,n-1} )/\beta_1$ and $\beta_1 > (1-c_{1,n})/(1+c_{1,n-1})$. 
\item \label{prty:thm_no04} There holds $x_{n-1}^* > x_1^*$ if and only if $\beta_2 > (1-c_{1,n-1})/(c_{1,n}+\beta_1)$ with $\beta_2 \in (0,1)$. There exists such a $\beta_2 \in (0,1)$ only if $\beta_1 > \sum_{i = 2}^{n-2} c_{1,i}$. 
%\item \label{prty:thm_no05} There holds $x_{n-1}^* > x_1^*$ if $\beta_1 > (1- c_{1,n-1} - \beta_2 c_{1,n} )/\beta_2$ and $\beta_2 > (1-c{1,n-1})/(1+c_{1,n})$. 
\item \label{prty:thm_no06} There holds $x_n^* < x_{n-1}^*$ if and only if $\beta_2 > \beta_1 c_{1,n} + c_{1,n-1} (c_{1,n} - 1)$ or equivalently $\beta_1 < (\beta_2 + c_{n-1} (1 - c_{1,n}))$ 
\end{enumerate} 
\end{theorem}

Note that the inequality in statement~\ref{prty:thm_no02} can be rewritten as $\beta_2 > (1-c_{1,n} - \beta_1 c_{1,n-1} )/\beta_1$ with $\beta_2 \in (0,1)$ which is satisfiable only if $\beta_1 > (1-c_{1,n})/(1+c_{1,n-1})$. Similarly, the inequality in statement~\ref{prty:thm_no04} is equivalent to $\beta_1 > (1- c_{1,n-1} - \beta_2 c_{1,n} )/\beta_2$ with $\beta_1 \in (0,1)$ which is satisfiable only if $\beta_2 > (1-c{1,n-1})/(1+c_{1,n})$.

We now interpret Statement~\ref{prty:thm_no02}, which we believe is the key result of the theorem. A similar conclusion can be drawn for Statement~\ref{prty:thm_no04} but we omit this due to spatial limitations. In order to make centre node $v_1$ lose social dominance, the dissent subject nodes $v_{n-1}$ and $v_{n}$ \emph{must adopt a cooperative strategy}. From their definitions, we can interpret $\beta_1$ as the trust given by $v_{n-1}$ to $v_n$ while $\beta_2$ is the trust given by $v_{n}$ to $v_{n-1}$. \emph{A necessary condition} for individual $v_n$ to have social power greater than centre node $v_1$ is that $\beta_2 > \sum_{i = 2}^{n-2} c_{1,i}$. This means that not only must $v_{n-1}$ trust $v_n$ sufficiently (as given by the inequality $\beta_1 > (1-c_{1,n})/(c_{1,n-1}+\beta_2)$), but individual $v_n$ must \emph{reciprocate} by ensuring that it trusts $v_{n-1}$ sufficiently ($\beta_1 > \sum_{i = 2}^{n-2} c_{1,i}$). Unless the two dissenting nodes build a cooperative and sufficiently strong bilateral relationship, centre node $v_1$ will remain socially dominant.

\subsection{Topology Variation~\ref{top:leader_group}: Leadership Group}

With $\beta_1 = c_{1,n+1} \in (0,1)$ and $\beta_2 = c_{n+1,1} \in (0,1)$, the following result is obtained 
\begin{theorem}\label{thm:leader_group}
For a social network with Topology Variation~\ref{top:leader_group}, with initial conditions $\vect{x}(s) \in \wt{\Delta}_{n+m}$, and described by the DeGroot-Friedkin model, the following statements are true:
\begin{enumerate}[label=(\roman*)]
\item \label{prty:thm_leader01} For all $\beta_1 \in (0,1)$ and for all $\beta_2 \in (0,1)$ there holds $x_i^* < x_1^*$ and $x_k^* < x_{n+1}^*$ for $i \in \{2, ..., n\}$ and $k \in \{n+2, ..., n+m\}$. 
\item \label{prty:thm_leader02} There holds $x_1^* < x_{n+1}^*$ or $x_1^* > x_{n+1}^*$ if and only if $\beta_2 < \beta_1$ or $\beta_2 > \beta_1$ respectively. If $\beta_1 = \beta_2$, then $x_1^* = x_{n+1}^*$.
\item \label{prty:thm_leader03} For $k \in \{n+2, ..., n+m\}$ there holds $x_k^* > x_1^*$ if and only if $c_{n+1,k}(\beta_1/\beta_2) > 1$. For $i \in \{ 2, ..., n\}$, holds $x_i^* > x_{n+1}^*$ if and only if $c_{1,i}(\beta_2/\beta_1) > 1$.
\end{enumerate}
\end{theorem}

Statement~\ref{prty:thm_leader02} shows that the ratio of $\beta_1/\beta_2$ determines whether centre node $v_1$ or centre node $v_{n+1}$ is socially dominant. Statement~\ref{prty:thm_leader03} delivers a surprising and interesting result on how leaders can \emph{cooperatively} protect themselves and maintain collective social dominance. Let $i \in \{2, ..., n\}$ and $k \in \{n+1, ..., n+m\}$. Consider from centre individual $v_1$'s point of view. While $x_i^* < x_1^*$ is guaranteed, in order to ensure that $v_1$ has greater social power than subject $v_k$ (i.e. subjects of centre individual $v_{n+1}$), individual $v_1$ must ensure that $c_{n+1,k}(\beta_1/\beta_2) < 1$. This inequality always holds, regardless of the value of $c_{n+1,k} < 1$, if $\beta_1 = \beta_2$. I.e. if the trust level $v_1$ accords to $v_{n+1}$ is equal to the trust level $v_{n+1}$ accords to $v_1$, regardless of the magnitude of $\beta_1 = \beta_2$, $v_1$ has greater social power than \emph{all} subject nodes \emph{including the subjects of $v_{n+1}$}. It can appear to be surprising because this holds even if $c_{n+1,k} >> \beta_1, \beta_2$. Yet such a result is intuitive if we consider $\beta_1/\beta_2$ as the ratio of the trust $v_1$ places on $v_{n+1}$ (and indirectly the trust $v_1$ places on subject $v_{k}$) versus the trust $v_{n+1}$ places on $v_1$ (and indirectly the trust $v_k$ places on $v_1$).

\section{Simulations}\label{sec:simulations}
In this section, we provide 2 short simulations to highlight some of our most interesting results. We do not provide comprehensive simulations for each Topology Variation due to spatial limitations. 

\begin{figure}
\begin{center}
\includegraphics[width=0.8\linewidth]{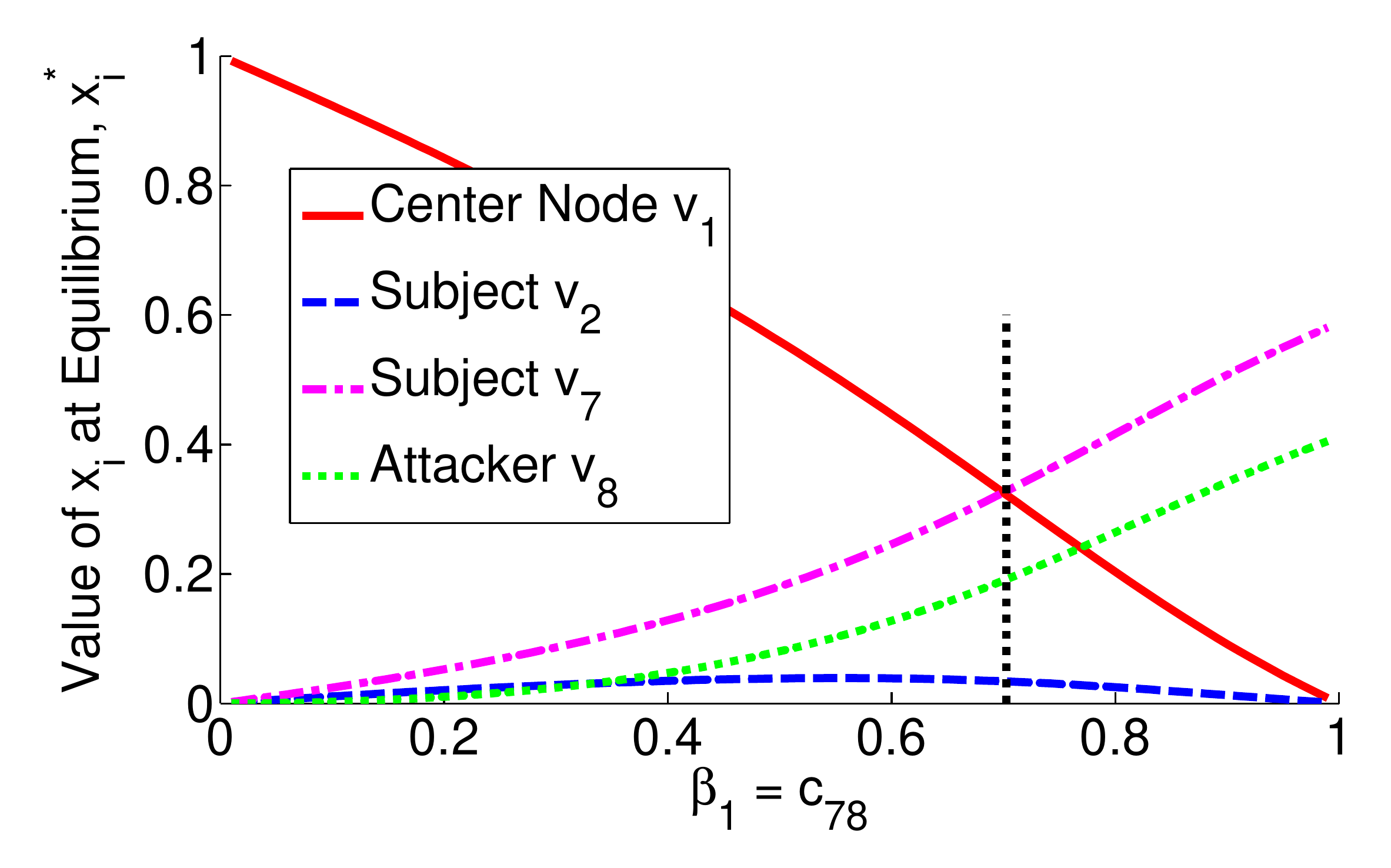}
\caption{Simulation of Topology Variation~\ref{top:single} with $n = 8$.}
\label{fig:SimTopVar1}
\end{center}
\end{figure}

Firstly, we simulate Topology Variation~\ref{top:single} as it is the fundamental strategy, with $n = 8$. The top row of the matrix $\mat{C}$ is given by $[0,\, 0.15,\, 0.15,\, 0.2,\, 0.05,\, 0.15,\, 0.3,\, 0]$. Figure~\ref{fig:SimTopVar1} shows the social power at equilibrium $x_i^*$, for selected individuals, as a function of $\beta = c_{78}$. Centre $v_1$ loses social dominance when $\beta > 0.7$ as stated in Theorem~\ref{thm:single}.

\begin{figure}
\begin{center}
\includegraphics[width=0.8\linewidth]{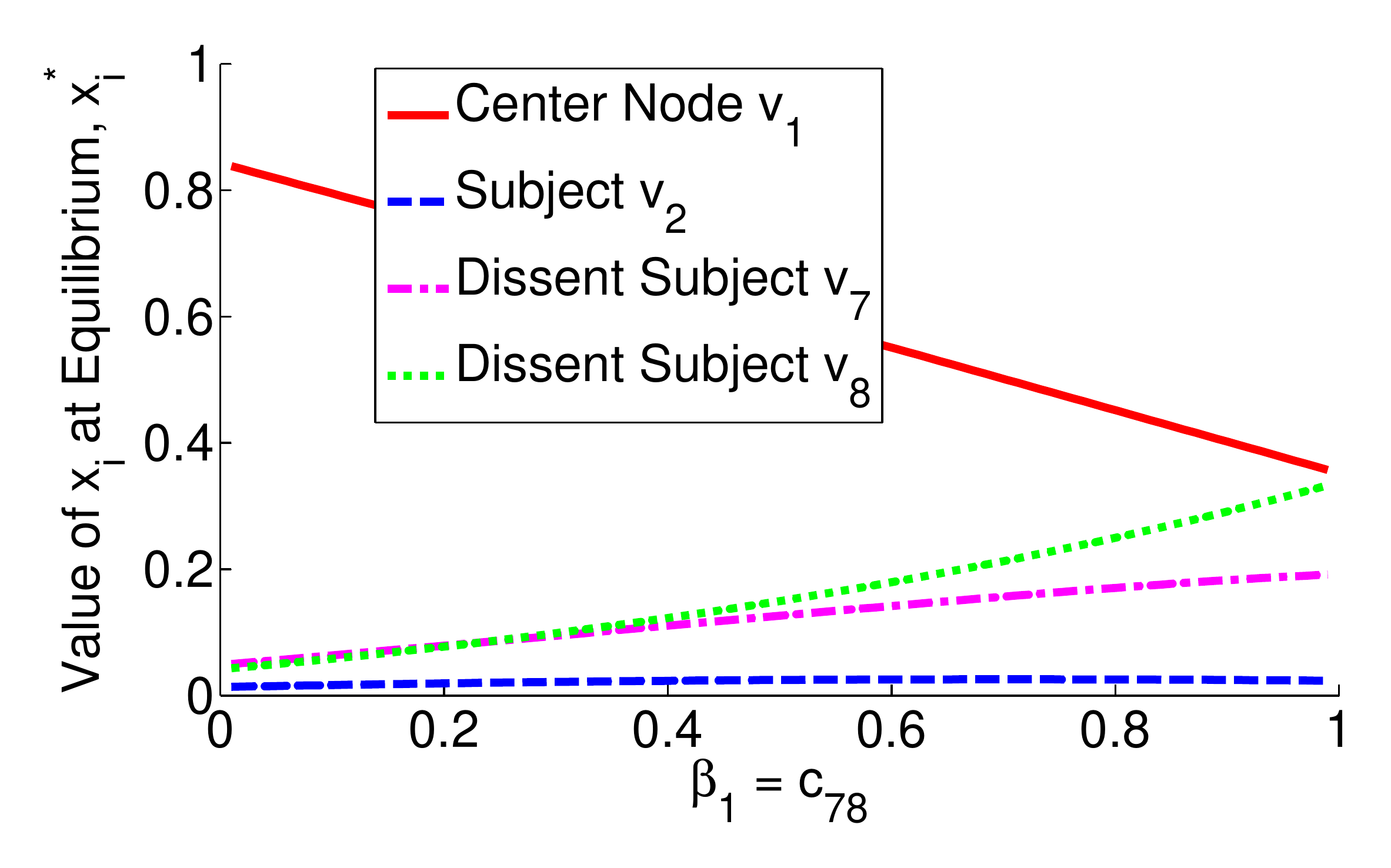}
\caption{Simulation of Topology Variation~\ref{top:no_attack} with $n = 8$, when $\beta_2 < \sum_{i = 2}^{n-2} c_{1,i}$.}
\label{fig:SimTopVar4_fail}
\end{center}
\end{figure}

\begin{figure}
\begin{center}
\includegraphics[width=0.8\linewidth]{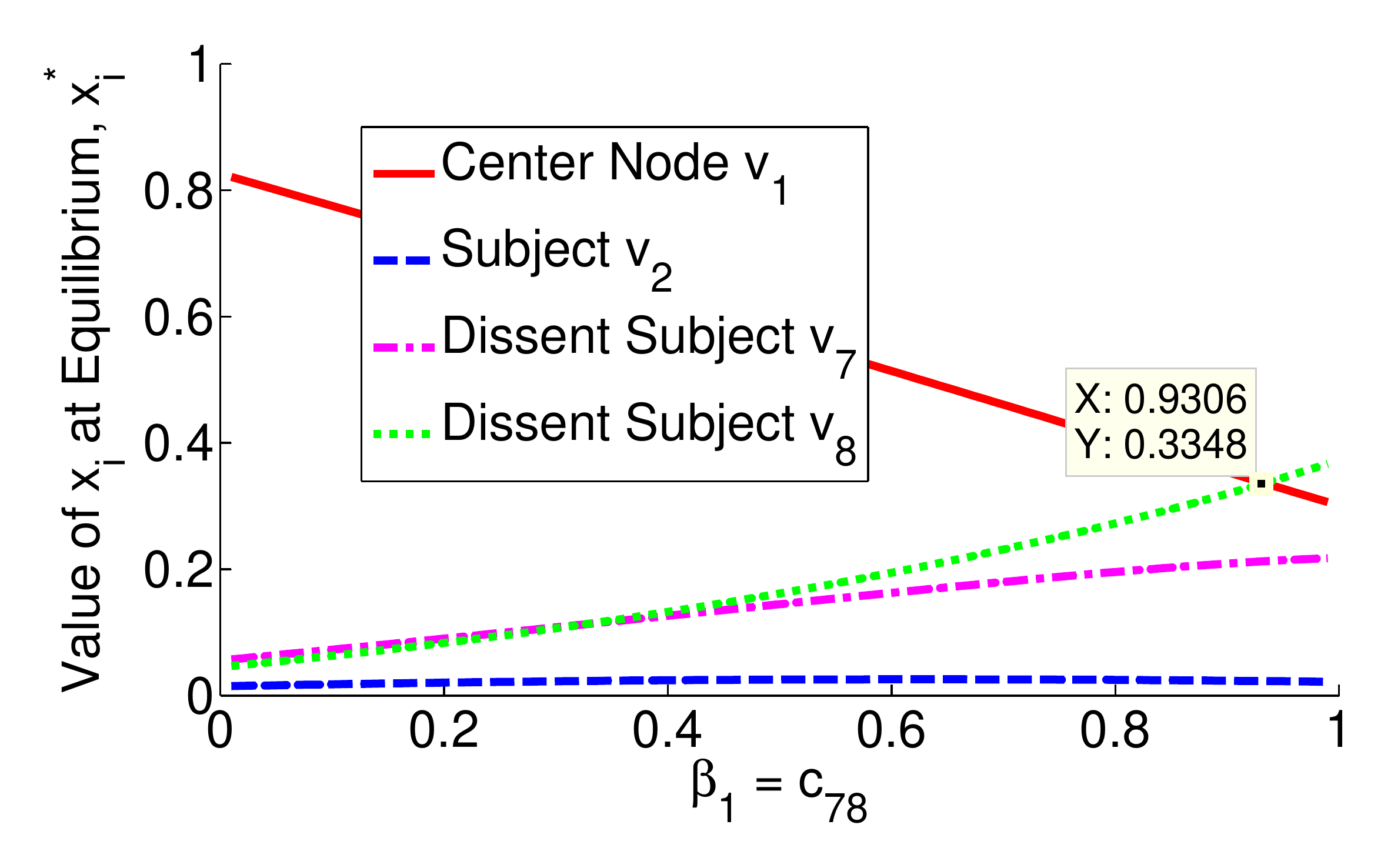}
\caption{Simulation of Topology Variation~\ref{top:no_attack} with $n = 8$, when $\beta_2 > \sum_{i = 2}^{n-2} c_{1,i}$.}
\label{fig:SimTopVar4_win}
\end{center}
\end{figure}

Next, we simulate Topology Variation~\ref{top:no_attack} to show the need for cooperation between two dissenting individuals in order to displace the centre node. The top row of $\mat{C}$ is $[0,\, 0.1,\, 0.1,\, 0.2,\, 0.05,\, 0.05,\, 0.2,\, 0.3]$. Figure~\ref{fig:SimTopVar4_fail} shows the social power at equilibrium $x_i^*$, for selected individuals as a function of $\beta_1 = c_{78}$ when $\beta_2 = 0.49$ (i.e. when $\beta_2 < \sum_{i = 2}^{n-2} c_{1,i}$). In accordance with Theorem~\ref{thm:no_attack}, Statement~\ref{prty:thm_no02}, dissent subject $v_8$ never achieves social power greater than centre $v_1$ because there does not exist a $\beta_1 \in (0,1)$ satisfying the required inequality. Figure~\ref{fig:SimTopVar4_win} shows the same simulation scenario but now with $\beta_2 = 0.55 > \sum_{i = 2}^{n-2} c_{1,i}$. In accordance with Statement~\ref{prty:thm_no02} of Theorem~\ref{thm:no_attack}, $x_8^* > x_1^*$ when $\beta_1 > 0.93$.

\section{Conclusions}\label{sec:conclusions}
Social networks with a star topology converge to an autocratic configuration, with the centre individual holding all the social power, as the number of issues discussed tend towards infinity. This paper proposed a number of different strategies, involving introduction of new individuals and/or new interpersonal relationships into the social network, in order to move social dominance from the centre individual to a subject individual. Necessary and sufficient conditions are developed, and interpretation of these conditions showed the strategies are sociologically intuitive. Numerous future directions exist. Firstly, we wish to generalise the results on uncoordinated attack and coordinated attack to arbitrary numbers of attacker nodes. Different leadership groups, and dissent topologies will also be explored. We also wish to investigate whether such straightforward strategies exist for more general topologies, and lastly we shall study strategies concerning social power for a subgroup of individuals.

\appendix

\section*{Proofs for Section~\ref{sec:main_result}}

In the following proofs, we make extensive use of Theorem~\ref{thm:DFmain}, and in particular Statement \ref{prty:thm_DFmain02}, which states that $x_i^* > x_j^* \Leftrightarrow \gamma_i > \gamma_j$ and that $x_i^* = x_j^* \Leftrightarrow \gamma_i = \gamma_j$.

\subsection{Theorem~\ref{thm:single} and Corollary~\ref{cor:single}}\label{app:pf_single}
The expression $\vect\gamma^\top = \vect\gamma^\top \mat C$, where $\mat{C}$ is given in \eqref{eq:C_singleattack}, allows us to obtain
\begin{subequations} \label{eq:c_cA}
\begin{align}
\gamma_1 &= \sum_{i = 2}^{n-2} \gamma_i + (1-\beta)\gamma_{n-1} \label{eq:c_cA_01} \\
\gamma_i & = c_{1i}\gamma_1,\quad\forall\, i \neq 1,n-1,n \label{eq:c_cA_02}\\
\gamma_{n-1} & = c_{1,n-1}\gamma_1 + \gamma_n \label{eq:c_cA_03}\\
\gamma_n & = \beta \gamma_{n-1} \label{eq:c_cA_04}
\end{align}
\end{subequations}
Statement~\ref{prty:thm_single01} is obtained from \eqref{eq:c_cA_02}, where we conclude $\gamma_i < \gamma_1$ because $c_{1i} < 1$ for all $i \neq 1, n-1, n$, and from \eqref{eq:c_cA_04}, which allows us to conclude that $\gamma_n < \gamma_{n-1}$ for all $\beta \in (0,1)$. For Statement~\ref{prty:thm_single02}, begin by substituting $\gamma_n$ from \eqref{eq:c_cA_04} into \eqref{eq:c_cA_03}, which yields $\gamma_{n-1} = c_{1,n-1}\gamma_1 + \beta \gamma_{n-1}$. This is rearranged to obtain
\begin{equation}\label{eq:disrupt_beta}
\gamma_1 = \frac{1-\beta}{c_{1,n-1}} \gamma_{n-1}
\end{equation}
Recalling that $0 < c_{1,n-1}$ and $0 < \beta < 1$, it follows that $\gamma_1 < \gamma_{n-1}$ if and only if $\beta > 1 - c_{1,n-1}$. Similarly, one can obtain that $\gamma_1 > \gamma_n$ if and only if $\beta < 1/(1+c_{1,n-1})$, which proves Statement~\ref{prty:thm_single03}. Corollary~\ref{cor:single} is a generalisation of Statement~\ref{prty:thm_single02} obtained by observing that $\text{argmin}_j (1-c_{1,j}) = \text{argmax}_j c_{1,j}$. \hfill $\square$

\subsection{Theorem~\ref{thm:two_attack} and Corollary~\ref{cor:two_attack}}\label{app:pf_twoattack}
For Topology Variation~\ref{top:two_attack}, the relative interaction matrix $\mat{C}$ is given by
\begin{equation}\label{eq:C_twoattackonesub}
\mat C(\beta) = 
\begin{bmatrix}
0 & c_{12} & c_{13} & \hdots & c_{1,n-1} & 0 & 0 \\
1 & 0 & 0 & \hdots & 0 & 0 & 0 \\
1 & 0 & 0 & \hdots & 0 & 0 & 0 \\
\vdots & \vdots & \vdots  & \ddots & \vdots & \vdots & \vdots \\
 1-(\beta_1+\beta_2) &  0 & 0 & \hdots & 0 & \beta_1 & \beta_2 \\
  0 & 0 & 0 & \hdots & 1 & 0 & 0 \\
    0 & 0 & 0 & \hdots & 1 & 0 & 0
\end{bmatrix}
\end{equation}
From $\vect{\gamma}^\top \mat{C} = \vect{\gamma}^\top$, we obtain
\begin{subequations} \label{eq:c_cA_2attack}
\begin{align}
\gamma_1 &= \sum_{i = 2}^{n-3} \gamma_i + (1-\beta_1-\beta_2)\gamma_{n-2} \label{eq:c_cA_2attack_01} \\
\gamma_i & = c_{1i}\gamma_1,\quad\forall\, i \neq 1,n-2, n-1,n \label{eq:c_cA_2attack_02}\\
\gamma_{n-2} & = c_{1,n-2}\gamma_1 + \gamma_{n-1} + \gamma_n \label{eq:c_cA_2attack_03}\\
\gamma_{n-1} & = \beta_1 \gamma_{n-2} \label{eq:c_cA_2attack_04} \\
\gamma_n & = \beta_2 \gamma_{n-2} \label{eq:c_cA_2attack_05}
\end{align}
\end{subequations}
Statement~\ref{prty:thm_two01} is obtained from \eqref{eq:c_cA_2attack_01} and \eqref{eq:c_cA_2attack_04} and \eqref{eq:c_cA_2attack_05}, using the same arguments as the proof for Theorem~\ref{thm:single}. Regarding Statement~\ref{prty:thm_two02}, substitute \eqref{eq:c_cA_2attack_04} and \eqref{eq:c_cA_2attack_05} into \eqref{eq:c_cA_2attack_03} and rearrange to obtain $\gamma_{n-2} = c_{1,n-2}\gamma_1/(1-\beta_1 - \beta_2)$. The statement is then straightforwardly obtained. For Statement~\ref{prty:thm_two03}, in regards to $\gamma_n$, substitute $\gamma_{n-2} = c_{1,n-2}\gamma_1/(1-\beta_1 - \beta_2)$ into the right hand side of \eqref{eq:c_cA_2attack_05} to obtain $\gamma_n = \beta_2 c_{1,n-2} \gamma_1/ (1 - \beta_1 - \beta_2)$. It is straightforward to verify that $\beta_2 > (1- \beta_1)/(1+c_{1,n-2})$ implies $\beta_2 c_{1,n-2}/ (1 - \beta_1 - \beta_2) > 1$, which in turn implies $\gamma_n > \gamma_1$. The inequality that ensures $\gamma_{n-1} > \gamma_1$ can be similarly found. Observe that  $1- \beta_1 < 1$, $1-\beta_2 < 1$ and $1 <  1+c_{1,n-2}$. There must also hold $\beta_1 + \beta_2 < 1$. This implies that for any value $c_{1,n-2}$, there always exist $\beta_1, \beta_2$ which ensures $\gamma_{n-1} > \gamma_1$ \emph{and} $\gamma_n > \gamma_1$. 
Regarding Statement~\ref{prty:thm_two04}, from \eqref{eq:c_cA_2attack_04} and \eqref{eq:c_cA_2attack_05}, we have $\gamma_{n-1}/\gamma_{n} = \beta_1/\beta_2$. The statement is then straightforwardly obtained.
Corollary~\ref{cor:two_attack} is a generalisation of Statement~\ref{prty:thm_two02} by observing that $\text{argmin}_j (1-c_{1,j}) = \text{argmax}_j c_{1,j}$.

\subsection{Theorem~\ref{thm:twotwo}}
The relative interaction matrix is given by
\begin{equation}
\mat{C}(\beta_1,\beta_2)=
\begin{bmatrix}
0 & c_{1,2} & \cdots & c_{1,n-3} & c_{1,n-2} & 0 & 0 \\
1 & 0 & \cdots & 0 & 0 & 0 & 0 \\
\vdots & \vdots  & \ddots  & \vdots  & \vdots  & \vdots  & \vdots  \\
1-\beta_1 & 0 & \cdots & 0 & 0 & \beta_1 & 0 \\
1-\beta_2 & 0 & \cdots & 0 & 0 & 0 & \beta_2 \\
0 & 0 & \cdots & 1 & 0 & 0 & 0 \\
0 & 0 & \cdots & 0 & 1 & 0 & 0 
\end{bmatrix}
\end{equation}
and the equation $\vect{\gamma}^\top \mat{C}= \vect{\gamma}^\top$ yields the following equalities:
\begin{subequations}\label{eq:v3relations}
\begin{align}
\gamma_1 &= (1-\beta_1)\gamma_{n-3} + (1-\beta_2)\gamma_{n-2} + \sum_{i=2}^{n-4} \gamma_i \label{eq:v3centre}\\
\gamma_i &= c_i\gamma_1, \;\;\; i\in\{2,\ldots,n-4\} \label{eq:v3subs}\\
\gamma_{n-3} &= c_{1,n-3}\gamma_{1} +\gamma_{n-1} \label{eq:v3sub1}\\
\gamma_{n-2} &= c_{1,n-2}\gamma_{1} +\gamma_{n} \label{eq:v3sub2}\\
\gamma_{n-1} &= \beta_1\gamma_{n-3} \label{eq:v3a1}\\
\gamma_{n} &= \beta_2\gamma_{n-2} \label{eq:v3a2}
\end{align}
\end{subequations}

From (\ref{eq:v3subs}), since $c_{1,i} \in (0,1)$ for all $i\in\{2,\ldots,n-4\}$, it follows that $\gamma_i<\gamma_0$ for all 
$i\in\{2,\ldots,n-4\}$. From (\ref{eq:v3a1}) and (\ref{eq:v3a2}), since $\beta_1,\beta_2\in (0,1)$, it follows that $\gamma_{n-1}<\gamma_{n-3}$ and $\gamma_{n}<\gamma_{n-2}$. Thus, statement~\ref{prty:thm_twotwo01} is true. 

From (\ref{eq:v3sub1}) and (\ref{eq:v3a1}), we have 
$\frac{ \gamma_{n-3} }{\gamma_1} = \frac{c_{1,n-3}} {1-\beta_1}$, 
which implies that $\gamma_1>\gamma_{n-3}$ if and only if 
$\beta_1 < 1-c_{1,n-3}$. Similarly, from (\ref{eq:v3sub2}) and (\ref{eq:v3a2}), we have $\gamma_1 >\gamma_{n-2}$ if and only if
$\beta_2 < 1-c_{1,n-2}$. It is then straightforward to conclude that for $i\in\{1,2\}$, if $\beta_i > 1 - c_{1,n-4+i}$, then $x_{n-4+i}^*> x_1^*$. Therefore, statement~\ref{prty:thm_twotwo02} is true.

From (\ref{eq:v3sub1}) and (\ref{eq:v3a1}), we have 
$\frac{\gamma_{n-1}}{\gamma_1} = \frac{\beta_1 c_{1,n-3} }{1-\beta_1}$.
It follows that $\gamma_{n-1}>\gamma_1$ if and only if 
$\beta_1>1/(1+c_{1,n-3})$. Similarly, from (\ref{eq:v3sub2}) and (\ref{eq:v3a2}), we have $\gamma_{n}>\gamma_1$ if and only if $\beta_2>1/(1+c_{1,n-2})$. Thus, statement~\ref{prty:thm_twotwo03} is true.

Since $\frac{\gamma_{n-3}}{\gamma_1} = \frac{c_{1,n-3}}{1-\beta_1}$ and $\frac{\gamma_{n-2}}{\gamma_1} = \frac{c_{1,n-2}}{1-\beta_2}$, it follows that $\gamma_{n-3}(1-\beta_1)/c_{1,n-3} = \gamma_{n-2}(1-\beta_2)/c_{1,n-2}$, which implies that $\frac{\gamma_{n-3}}{\gamma_{n-2}} = \frac{c_{1,n-3}(1-\beta_2)}{c_{1,n-2}(1-\beta_1)}$.
Then, $\gamma_{n-3} > \gamma_{n-2}$ if and only if $\frac{1-\beta_2}{1-\beta_1}>\frac{c_{1,n-2}}{c_{1,n-3}}$. Therefore, statement~\ref{prty:thm_twotwo04} is true.
\hfill$\qed$

\subsection{Theorem~\ref{thm:no_attack}}\label{app:pf_noattack}
For Topology Variation~\ref{top:no_attack}, the relative interaction matrix $\mat{C}$ is expressed as
\begin{equation}
\mat C(\beta_1, \beta_2) = 
\begin{bmatrix}
0 & c_{12} & c_{13} & \hdots & c_{1,n-1} & c_{1,n} \\
1 & 0 & 0 & \hdots & 0 & 0 \\
\vdots & \vdots & \vdots  & \ddots & \vdots & \vdots \\
 1-\beta_1 &  0 & 0 & \hdots & 0 & \beta_1 \\
 1- \beta_2 & 0 & 0 & 0 & \beta_2 & 0
\end{bmatrix}
\end{equation}
where $\beta_1 = c_{n-1,n}$ and $\beta_2 = c_{n,n-1}$. The expression $\vect{\gamma}^\top \mat{C} = \vect{\gamma}^\top$ yields the following equalities
\begin{subequations} \label{eq:c_cA_no}
\begin{align}
\gamma_1 &= \sum_{i =2 }^{n-2} \gamma_i + (1-\beta_1)\gamma_{n-1} + (1-\beta_2) \gamma_n \label{eq:c_cA_noattack_01} \\
\gamma_i & = c_{1i}\gamma_1,\quad\forall\, i \neq 1, n-1,n \label{eq:c_cA_noattack_02}\\
\gamma_{n-1} & = c_{1,n-1}\gamma_1 + \beta_2\gamma_n \label{eq:c_cA_noattack_03}\\
\gamma_{n} & = c_{1,n}\gamma_1 + \beta_1 \gamma_{n-1} \label{eq:c_cA_noattack_04}
\end{align}
\end{subequations}
Again, Statement~\ref{prty:thm_no01} is obtained trivially from \eqref{eq:c_cA_noattack_02}. Substitute \eqref{eq:c_cA_noattack_03} into \eqref{eq:c_cA_noattack_04} and rearrange for $\gamma_n$ to obtain
\begin{equation}
\gamma_n = \left( \frac{c_{1,n} + \beta_1 c_{1,n-1} }{ 1-\beta_1\beta_2 } \right) \gamma_1
\end{equation}
and it follows that $\gamma_n > \gamma_1$ is implied by 
\begin{align}
c_{1,n}+ \beta_1 c_{1, n-1} & > 1 - \beta_1 \beta_2 \\
\beta_1 > \frac{1-c_{1,n}}{c_{1,n-1} + \beta_2 } \label{eq:cond_noattack_b1} \\
\beta_2 > \frac{1 - c_{1,n} - c_{1,n-1}\beta_1 }{\beta_1 } \label{eq:cond_noattack_b2}
\end{align}
Consider \eqref{eq:cond_noattack_b1}. Observe that $(1-c_{1,n})/(c_{1,n-1} + \beta_2) \geq 1 \Leftrightarrow 1 - c_{1,n} - c_{1,n-1} \geq \beta_2 \Leftrightarrow \sum_{i = 2}^{n-2} c_{1,i} \geq \beta_2$. Recalling that $\beta_1 \in (0,1)$, we conclude $\gamma_n > \gamma_1$ is possible only if $\beta_2 > \sum_{i = 2}^{n-2} c_{1,i}$. Alternatively, one can consider \eqref{eq:cond_noattack_b2} and similarly derive that $\gamma_n > \gamma_1$ if $\beta_2 > (1-c_{1,n} - \beta_1 c_{1,n-1})/\beta_1$ and $\beta_1 > (1-c_{1,n})/(1+c_{1,n-1})$. The inequality conditions for ensuring $\gamma_{n-1} > \gamma_1$ are also derived in similar manner and omitted due to spatial limitations.

\section{Proofs for Section~\ref{sec:leadership_groups}}

\subsection{Theorem~\ref{thm:leader_group}}
The relative interaction matrix for Topology Variation~\ref{top:leader_group} is given by
\begin{align}
& \mat C(\beta_1, \beta_2) \nonumber \\
& = 
\begin{bmatrix}
0 & c_{12} & c_{13} & \hdots & \beta_1 & 0 & \hdots & 0 \\
1 & 0 & 0 & \hdots & 0 & 0 & \hdots & 0 \\
\vdots & \vdots & \vdots  & \ddots & \vdots & \vdots & \ddots & \vdots \\
1 & 0 & 0 & \hdots & 0 & 0 & \hdots & 0 \\
 \beta_2 &  0 & 0 & \hdots & 0 & c_{n+1,n+2} & \hdots & c_{n+1,n} \\
0 & 0 & 0 & \hdots & 1 & 0 & \hdots & 0 \\
\vdots & \vdots & \vdots  & \ddots & \vdots & \vdots & \ddots & \vdots \\
0 & 0 & 0 & \hdots & 1 & 0 & \hdots & 0
\end{bmatrix}
\end{align}
And the expression $\vect{\gamma}^\top \mat{C} = \vect{\gamma}^\top$ yields the following equalities
\begin{subequations} \label{eq:c_cA_leader}
\begin{align}
\gamma_1 &= \sum_{ 1 < i \leq n  } \gamma_i + \beta_2\gamma_{n+1} \label{eq:c_cA_leader_01} \\
\gamma_i & = c_{1,i}\gamma_1,\quad\forall\, i \in \{2, ..., n\} \label{eq:c_cA_leader_02}\\
\gamma_{n+1} & = \sum_{ n + 1 < i \leq n+m  } \gamma_i  + \beta_1\gamma_1 \label{eq:c_cA_leader_03}\\
\gamma_{i} & = c_{n+1,i} \gamma_{n+1},\quad\forall\, i \in \{n+2, ..., n+m\}  \label{eq:c_cA_leader_04}
\end{align}
\end{subequations}
Statement~\ref{prty:thm_leader01} is obtained trivially from \eqref{eq:c_cA_leader_02} and \eqref{eq:c_cA_leader_04}. In regards to Statement~\ref{prty:thm_leader02}, first substitute \eqref{eq:c_cA_leader_02} into \eqref{eq:c_cA_leader_01} to obtain $\gamma_1 = \beta_2 \gamma_{n+1} + \sum_{ 1 < i \leq n  } c_{1,i} \gamma_1$ which is rearranged to yield $\gamma_1 (1 - \sum_{ 1 < i \leq n  } c_{1,i}) = \beta_2 \gamma_{n+1}$ which is equivalent to $\beta_1\gamma_1  = \beta_2 \gamma_{n+1}$ because $1 - \sum_{ 1 < i \leq n  } c_{1,i} = \beta_1$. Statement~\ref{prty:thm_leader03} is obtained by substituting $\gamma_1 = \beta_2 \gamma_{n+1}/ \beta_1$ into \eqref{eq:c_cA_leader_04}.

\bibliographystyle{IEEEtran}
\bibliography{MYE_ANU,ji}

%\end{IEEEbiography}
%% insert where needed to balance the two columns on the last page with
%%\newpage


%\vfill


\end{document}